\documentclass[prl,aps,twocolumn]{revtex4}
\usepackage{epsfig}
\usepackage{amsmath}
\usepackage{bm}

\newcommand{\beq}{\begin{equation}}
\newcommand{\eeq}{\end{equation}}
\newcommand{\bea}{\begin{eqnarray}}
\newcommand{\eea}{\end{eqnarray}}
\newcommand{\bei}{\begin{itemize}}
\newcommand{\eei}{\end{itemize}}

\newcommand{\etal}{{\em et al.}}

\newcommand{\mui}{\vec{\mu}(\vec{r}_i)}
\newcommand{\muj}{\vec{\mu}(\vec{r}_j)}

\def\tit#1#2#3#4#5{{#1}{\bf #2}, #3 (#4)}

\def\epl{Europhys.\ Lett.\ }

\def\prl{Phys.\ Rev.\ Lett.\ }

\def\prb{Phys.\ Rev.\ B\ }

\def\sci{Science\ }

\def\jsp{J.\ Stat.\ Phys.\ }

\def\jmmm{J. Mag.\ and Mag.\ Mat.\ }

\def\suhe#1{{\bf #1}:}

\begin{document}

\title{Artificial square ice and related dipolar nanoarrays}

\author{G. M\"{o}ller$^1$ and R. Moessner$^2$}
\affiliation{$^1$Laboratoire de Physique Th\'eorique et Mod\`eles Statistiques, CNRS-UMR8626, 91406 Orsay, France}
\affiliation{$^2$Laboratoire de Physique Th\'eorique de l'Ecole Normale
Sup\'erieure, CNRS-UMR8549, Paris, France}

\begin{abstract}
We study a frustrated dipolar array recently manufactured lithographically by
Wang {\em et al.}\ [Nature {\bf 439}, 303 (2006)] in order to realize the square ice
model in an artificial structure. We discuss models for thermodynamics and
dynamics of this system. We show that an ice regime can be stabilized by small
changes in the array geometry; a different magnetic state, kagome ice, can
similarly be constructed. At low temperatures, the square ice regime is
terminated by a thermodynamic ordering transition, which can be chosen to be
ferro- or antiferromagnetic. We show that the arrays do not fully
equilibrate experimentally, and
identify a likely dynamical bottleneck.
\end{abstract}
\date{April 4, 2006}

\maketitle

\suhe{Introduction}
The ability to manipulate constituent degrees of freedom of condensed
matter systems and their interactions is fundamental to attempts
to advance our understanding of the variety of phenomena presented to us by
nature. For a long time this has been achieved by utilizing the
combinatorial richness of the periodic table of elements to construct
different chemical compounds. 

A more recent option is to use the tools of nanotechnology to
custom-tailor degrees of freedom which can be assembled in a highly
controlled manner; e.g.\ this has been proposed for realizing a
topologically protected quantum computer using Josephson Junction
arrays \cite{ioffenature}.  Submicron superconducting rings have also
been used to provide effective spin-1/2 degrees of freedom
\cite{reichring}.

\begin{figure}
\includegraphics*[width=0.91\columnwidth]{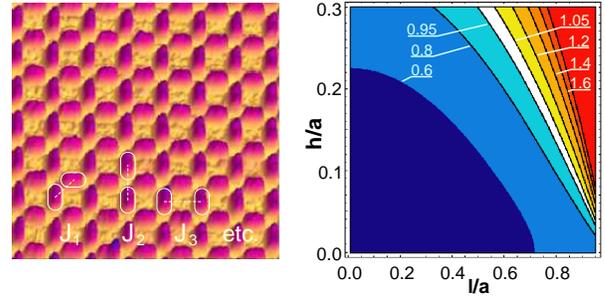}
  \caption{(color online) Left: Atomic force microscope image of an
  array studied in Ref.~\onlinecite{wangschiffer}. The islands have length
  $l=220$nm, width 80nm and thickness 25nm. Right: Map of the ratio
  $J_2/J_1$ of the second to the first nearest neighbor interactions
  (highlighted in the left part)
  for different values of lattice constant, $a$, and sublattice height
  offset, $h$.  
  In the white zone, $|J_2/J_1-1|<5\%$. In the left (blue)
  region, the ordered state is antiferromagnetic, whereas it is
  ferromagnetic in the right (yellow-red) area.  }
\label{fig:arraymfm}
\label{fig:mapJ2J1}
\end{figure}

Very recently, Wang and collaborators have used lithographic
techniques 
to create a periodic two-dimensional array of single-domain
submicron
ferromagnetic islands \cite{wangschiffer}, depicted in
Fig.~\ref{fig:arraymfm}.  
This design approach takes advantage of well-established lithographic
techniques and enables reading of the state of the system with local probes,
such as magnetic force microscopes, to image the state of single constituent
magnetic islands \cite{nanoreview}.

The first aim of this study is to
assemble a system that realizes the square ice model. 
This is an attractive target model because of its long and
distinguished history during which algebraic correlations and a finite
entropy at zero temperature have been established, as well as
connections to exact solutions,  quantum magnetism,
unusual dynamics and gauge theories \cite{liebice}.

The pioneering study by Wang \etal\ raises a number of important
questions which we try to address here. Firstly, what are appropriate
models for the arrays' thermodynamics and dynamics?  Secondly, what
other systems can one hope to build with these techniques? And
thirdly, what are interesting directions in which further developments
would be desirable?

In particular, the question of whether a dipolar system with long-range
interactions can be modelled by the short-range ice model is rather
similar to the one posed in the case of (three-dimensional) dipolar
spin ice,  
where it was found that a nearest-neighbor description was
surprisingly accurate for a range of properties, such as the
low-temperature (Pauling)
entropy \cite{stevenews,siddice,spinicereview,melkoice,projequiv}.

In this paper we show that an analogous equivalence between ice states
and the ground states of two-dimensional dipoles on the links of the
square lattice is more delicate. However, it can be established via a
route quite different from the three-dimensional case, namely by (a)
placing the dipoles pointing in different directions onto slightly
different heights and (b) manufacturing the dipoles as elongated as
possible. It turns out to be easier to realize kagome ice in a
dipolar array, as the requisite symmetry is compatible with embedding
in a plane.

As a byproduct, the low-temperature antiferromagnetic (in ice
language: antiferroelectric) instability of the original model can be
designed to be replaced by a ferromagnetic one.  However, the
experiments observe no ordering transition, implying that the array
does not fully equilibrate. We are thus led to study a
phenomenological model for its dynamics: zero-temperature (`greedy')
stochastic dynamics subject to an energy barrier for spin flips.  This
reproduces the experimental measurements semi-quantitatively.
Such dynamics are insufficient to
anneal out isolated defects violating the ice rules even for strong
interactions, thus preventing the establishment of an ice---or 
indeed an ordered---configuration.

In the remainder of this paper, we first analyze---by a mean-field
theory and using Monte-Carlo simulations---the equilibrium
statistical mechanics of arrays designed to mimic square and kagome
ice. We then turn our attention to the dynamics of the arrays studied
in Ref.~\onlinecite{wangschiffer} and present our results for the
local correlations. We conclude with
a brief discussion of disorder and an outlook.

\noindent\suhe{Dipoles on the links of a square lattice}
The interactions between the magnetic islands are dipolar, and
therefore essentially a  geometric property of the
array (lattice constant: $a$), described by the Hamiltonian
\bea
H=\iint d\vec{r}_i d\vec{r}_j 
\frac{
\mui\cdot\muj-3[\mui\cdot\hat{r}_{ij}][\muj\cdot\hat{r}_{ij}]}{r_{ij}^3},
\label{eq:hamilneedle}
\eea
where $\vec{r}_{ij}=\vec{r}_{i}-\vec{r}_{j}$, and $\mui$ is the dipole
moment at $\vec{r}_i$, which points along the link. 
We treat the dipoles either as points
($l/a\to0$) or as uniform, monodomain thin needles of finite length $l$.

Can such an arrangement be used to access square ice physics? In other
words, is there a (low-temperature) regime for such a dipolar system,
where configurations obeying the ice rule are overwhelmingly present
with approximately equal weights? 

This is, by construction, the case for the `Ising-ice model', in which
the four islands emanating from a given vertex 
of the square lattice interact
equally and antiferromagnetically, so that in the ground state two
dipoles point into each vertex, and two out. These are precisely the
ice states.
The Fourier transform of this interaction has two branches (as there
are two sites in the unit cell, one for each link direction of the
square lattice), which have eigenvalues
\bea
\tilde{J}_l(q)=0 \ \ ; 
\tilde{J}_u(q)=\,2J\left(\sin^2\frac{q_x}{2}+\sin^2\frac{q_y}{2}\right).
\eea
It is the flatness of the lower branch that indicates the frustration
of the ice model. As all the ice states are linear combinations of the
flat-band eigenvectors only, a sufficient condition for
Eq.~\ref{eq:hamilneedle} to yield an ice regime is that its lower
branch be flat and share the eigenvectors of the Ising-ice model
\cite{projequiv}, at least approximately.

By contrast, for a nearest-neighbor interaction only, one obtains a
symmetric pair of bands:
\bea
\tilde{J}_{l,u}(q)=\mp 2J \left|\sin\frac{q_x}{2}\sin\frac{q_y}{2}\right| .
\eea
\begin{figure}
  \includegraphics[width=0.48\columnwidth]{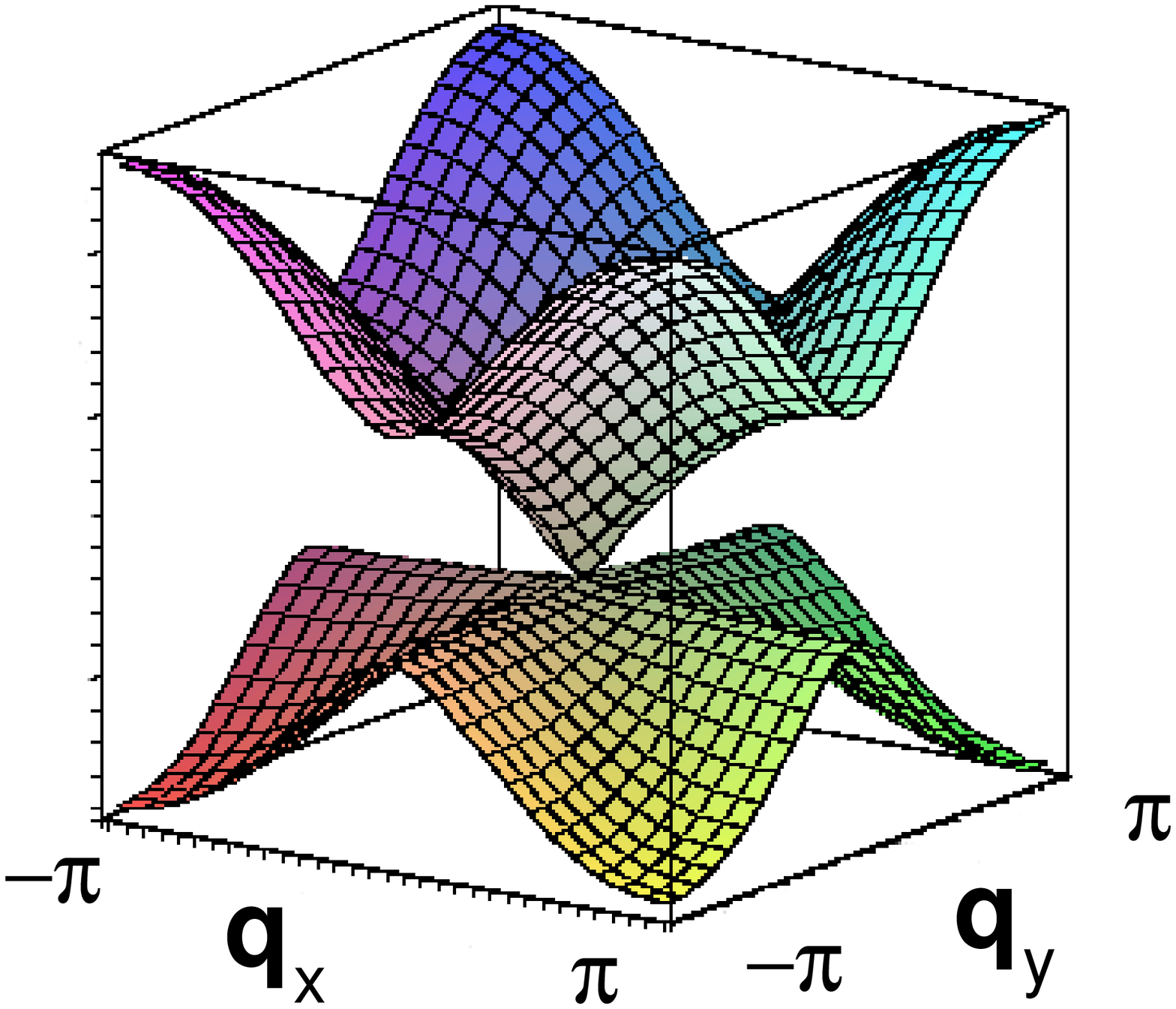}
\includegraphics[width=0.48\columnwidth]{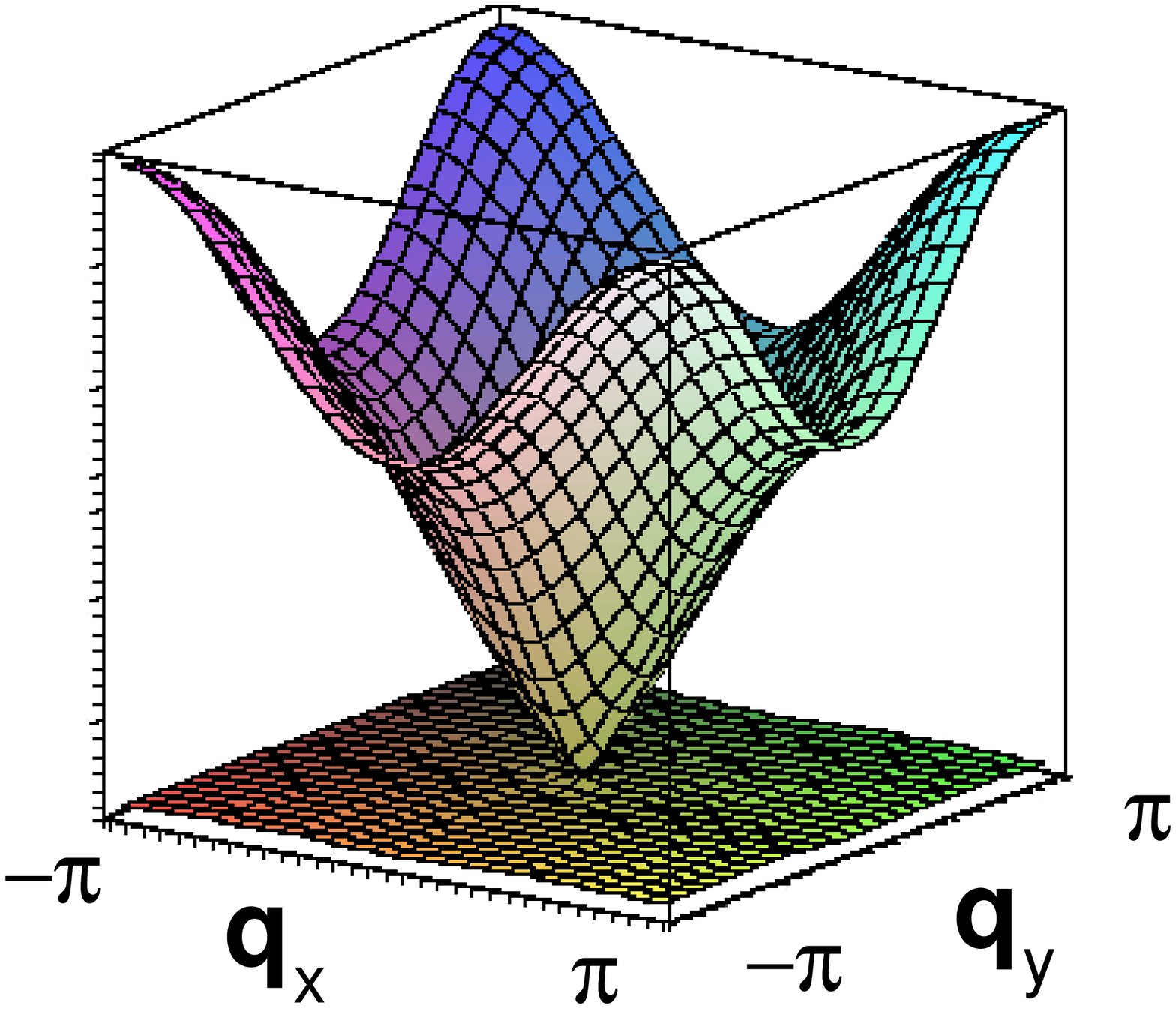}\\ 
\caption{(color online) Left: Spectrum for artificial ice 
(cut off after 10$a$, for point-like dipoles).  Right: same for
$l/a=0.7$ and $h/a=0.207$: the lower band becomes almost flat, 
(less than $1.5\%$ of the total bandwidth).
}
\label{fig:articespectrum}
\end{figure}

The spectrum for artificial ice (Fig.~\ref{fig:articespectrum}) 
lies in between, with a
substantial dispersion to the flat `ice' band, of about a
third
of the total bandwidth.  Thus, there is a thermodynamic transition to
antiferromagnetic order, which in our Monte-Carlo simulations occurs
at temperature $T_{\text{af}}=1.70(5)J_1$ (where $J_n$ is the strength of
the $n^{\text{th}}$-neighbor interaction, see 
examples in Fig. \ref{fig:arraymfm}).

\begin{figure}
\includegraphics[width=0.98\columnwidth]{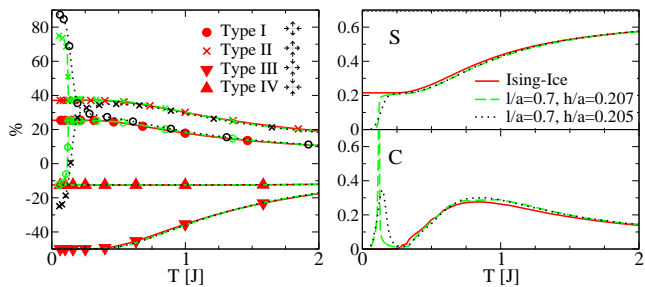} 
\caption{(color online) Ising-ice vs.\ dipolar arrays: 
frequency of vertex types (\% deviation from random distribution for
single vertex, which is $\frac18$, $\frac14$, $\frac12$ and $\frac18$
for Types I-IV, respectively), entropy (S), and heat capacity (C) from
Monte-Carlo simulations; $x$-axes were scaled for high-$T$
asymptotics to coincide. The fraction of non-ice-rule vertices is
below 1\% for $T<0.42 J$ for all depicted systems. At low $T$, the
ice regime, which widens with increasing $l/a$, is terminated by
ferro- (dashed) or antiferromagnetic (dots) order for different $h$.}
\label{fig:MCSquareIce}
\end{figure}

The principal reason for the dispersion of the lower band is the
inequivalence---as in the F-model---of the six vertices of the ice
model, which fall into two groups. One pair (labelled Type I in
Ref.~\onlinecite{wangschiffer}, see Fig.~\ref{fig:MCSquareIce}) has
zero total magnetic moment, while the others (Type II) have a net
moment along a diagonal and are higher in energy.
This inequivalence results from the fact that, unlike the case of
a tetrahedron in $d=3$, the six bonds between the four islands
belonging to a vertex are not all equivalent. 

However, this can be remedied by
introducing a height displacement $h$ between magnetic islands
pointing in the $x$- and $y$-directions.

The ratio $J_2/J_1$ of the two inequivalent bond
energies (Eq.~\ref{eq:hamilneedle}) 
is shown in Fig.~\ref{fig:mapJ2J1}
as a function of $h/a$ and $l/a$.  There is a set of choices for these
parameters such that the interaction energies are approximately equal.
For point-like dipoles ($l/a\rightarrow0$), this value is
\bea
h_{\text{ice}}/a=\sqrt{\left({3/8}\right)^{2/5}-1/2}
\approx0.419\,,
\eea 
and taking into account the finite extension of the dipoles lowers the
required height offset. For instance, for $l/a=0.7$,
$h_{\text{ice}}/a\approx0.207$. In principle, for $1-l/a\equiv\epsilon\to 0$,
$h_{\text{ice}}/a\sim\sqrt{2}\epsilon\to0$.  However, in this 
ideal limit the effects of
disorder, finite transverse width and a possible internal structure
of the dipoles will all play a role.

Having fixed the short-distance trouble by introducing the
modulation in height, the question remains what happens to the
long-distance part of the dipolar interaction, which in $d=3$,
amazingly, turned out to leave the ice regime intact
\cite{siddice,melkoice}. 
However, the mechanism responsible for this equivalence in $d=3$ \cite{projequiv}
is not operational in $d=2$, as it requires the dimensionality of the dipolar 
interaction to coincide with that of the underlying lattice.
Here, however, we have a $d=3$, $1/r^3$ dipolar interaction in a $d=2$
array.  Nonetheless, the present situation is relatively
benign, as the Fourier
sum of a $1/r^3$ interaction in $d=2$ is absolutely convergent 
(obviating the need for an Ewald sum).

Further neighbor terms can be suppressed parametrically 
in the ideal limit of $l/a\to 1$.  The ratio 
$J_{n\geq3}/J_{1,2}$
vanishes as $\epsilon\to 0$, thus yielding the ideal Ising-ice model.
As $l/a$ is reduced, the flat band initially acquires only a small
dispersion. To demonstrate this, in Fig.~\ref{fig:articespectrum} we
have plotted the mode spectrum for $l/a=0.7$, which corresponds to the
$a=320$nm sample \cite{wangschiffer}. The overlap of its eigenvectors
with those of the Ising-ice model differs from 1 by less than $0.1$\%
over the entire Brillouin zone.

This demonstrates that an ice regime can be obtained by this route.
Our Monte-Carlo simulations on the dipolar and the Ising-ice model for
$l/a=0.7$ (Fig.~\ref{fig:MCSquareIce}) bear out this statement: the intermediate
ice regime is terminated at high $T$ by thermally activated defects
violating the ice rules, and at low $T$ by an ordering
transition. Choosing $h$ on either side of the optimal value
$h_{\text{ice}}$, this transition is ferro/antiferromagnetic, respectively,
and perhaps even to a more complicated state very close to $h_{\text{ice}}$.

\noindent\suhe{Kagome ice}
The ground states of antiferromagnetic Ising spins on the kagome
lattice define what is known as kagome ice, with the `ice rules'
requiring each triangle to have one spin pointing in and two out, or vice
versa \cite{willsice}. Since the sites of the kagome lattice correspond to
the links of the honeycomb lattice, one can pose the question whether
a dipolar array forming a honeycomb lattice will display a kagome ice
regime.

The case of kagome ice has the advantage that the three bonds of the
triangle are equivalent, unlike the six
bonds of the square. This means that the nearest-neighbor Hamiltonian
does not require any fine-tuning through a height
offset. Furthermore, using the above limit of $\epsilon\to0$, we can again
parametrically suppress the importance of further-neighbor
interactions, and hence obtain a representation of kagome ice. The
vestiges of the further-neighbor terms will again give rise to an
ordering transition, terminating the ice phase on its low-$T$ side. We
note that kagome ice is a phase distinct from square ice in that its
long-wavelength theory is different---its correlations are 
not algebraic, but exponentially short-ranged
even at zero temperature.

\noindent\suhe{Dynamics and annealing}
Given the impossibility of thermally
equilibrating the array, the authors of Ref.~\onlinecite{wangschiffer} used
a rotating magnetic field $B_{\text{ext}}$, gradually stepped down, 
to speed up the
dynamics \cite{cowburn}. The question whether such an `algorithm'
can be efficiently used to find the ground state of a
system has been discussed in the context of spin glasses
\cite{zimanyi}. (However, in the case of an ice regime, 
the question is somewhat 
simpler, namely whether it is possible to find one of
{\em exponentially} many ice configurations.)

For our model of dipolar needles, the energy scales for the arrays
studied in Ref.~\onlinecite{wangschiffer} are:
\bea
\text{`Zeeman'\ energy}:&& 
|\mu\cdot B_{\text{ext}}|\leq 2.6\times 10^6 \mathrm{K}
\label{eq:zeeman}\\
\text{`Exchange'}:&&
3.6\times10^3 \mathrm{K} 
\leq J_{1}\leq 
1.1\times 10^5\mathrm{K} 
\nonumber
\eea
Antiferromagnetic long-range order should thus be present at room
temperature.

We consider a phenomenological model for the dynamics, motivated by
the experimental protocol.  Firstly, with the experimental temperatures well
below the interaction strengths, we use a form of zero-temperature
Monte-Carlo dynamics. Secondly, we note that vertices violating the ice rules are
experimentally present throughout. As such defect vertices
can be removed using single spin flips only, we impose a constraint on
our single-spin-flip dynamics: for a flip
to be accepted the energy gain must be above a threshold $\theta$.
Thirdly, the rotating field is modeled by a field of random
orientation.
 
\begin{figure}
\includegraphics*[width=0.95\columnwidth]{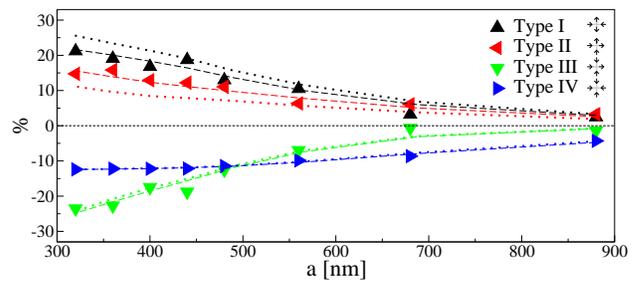}
\caption{(color online) 
Frequency of vertex types (as in Fig.~\ref{fig:MCSquareIce}).
Interaction energies scale approximately as $1/a^3$, a factor of 30
between the extremal points. Experiments (symbols) are shown against
dynamics simulations for needle dipoles (Eq.~\ref{eq:hamilneedle},
dotted), and for increased $J_2$ (dashed, see text).}
\label{fig:greedyexpt}
\end{figure}

This model has two free parameters: the threshold $\theta$ and the speed with
which the field is ramped down $\kappa$.  We fix these parameters to be the
same for all arrays, and fit them to obtain the best agreement with
the experimental measurements of the local correlations (longer-range
correlations in the experiment are very weak)
\cite{wangschiffer}. The best fit is
obtained for $\theta=2.3\times 10^5$ Kelvin and $\kappa=87$ Tesla$^{-1}$ 
attempted flips per spin during ramp-down (Fig.~\ref{fig:greedyexpt}).

This algorithm gives semi-quantitative agreement with experiment
over a range of interaction energies differing by a factor of 30
(Eq.~\ref{eq:zeeman}).  
However, we systematically overestimate the
frequency of Type I vertices compared to those of Type II (as do other
algorithms we have studied). 
This appears to be due to our needle model (Eq.~\ref{eq:hamilneedle})
overestimating the ratio $J_1/J_2$.  Denoting ${\cal
E}_{ij}=\frac{E_i-E_I}{E_j-E_I}$, where $E_i$ is the energy of an
isolated Type $i$ vertex, Ref.~\onlinecite{wangschiffer} finds from a
finite-element simulation: ${\cal E}_{32}>2$, ${\cal E}_{42}>6$ for
$l/a=0.7$. If we reduce the value of $|J_1-J_2|$ from
Eq.~\ref{eq:hamilneedle} by 30\%, our values grow from ${\cal
E}_{32}=1.7$, ${\cal E}_{42}=4.9$, to 2.25 and 7, respectively. The
resulting fit (Fig.~\ref{fig:greedyexpt}, dashed lines) appears
noise-limited \cite{fn-reduceice}.

Even for the strongest interactions, about 25\% (non-ice-rule)
defects, Type III vertices, persist.  Whereas it is easy to remove
pairs of appropriately oriented neighboring defects by flipping the
spin which joins them, such an annihilation process occurs with low
probability once these defects are sparse: they first need to diffuse
around until they encounter a partner.

\noindent\suhe{Disorder}
Our dynamical model does not take into account disorder, which is
expected to have substantial influence on the dynamical behavior even
of single islands \cite{nanoreview}.  
Thus, the
good agreement of our model with the experiment might be 
due to its correct
reproduction of the dynamical bottleneck, and not the detailed
microscopic dynamics: the inability of the defects to `find' one
another may e.g. simply be due to their becoming pinned.

Disorder also impacts the ice regime thermodynamically, 
as the size of the leading
perturbation sets the scale for its termination at low temperature. 
Especially for fine-tuned $h\sim h_{\text{ice}}$, disorder might dominate
(over $J_1\neq J_2$ and further-neighbor interactions), by selecting
some ice configurations over others; strong disorder might even lead
to the presence of defects at any temperature.

To push the analysis further in this direction, experimental input
would be desirable. What is the variance of the islands' geometrical
properties?  Are there some islands that freeze at much higher fields
than others? Are defects usually located at the same positions, and
what is their spatial distribution? How do correlations evolve during
the ramp-down of the external field?

\noindent\suhe{Summary and outlook}
We have presented models for the dynamics and thermodynamics of
frustrated dipolar arrays, including ways of stabilizing ice
regimes.  Perhaps the most interesting direction of further study
involves their dynamics, in the presence of varying degrees of
disorder. In particular, can other protocols \cite{zimanyi}, e.g.\ involving the
use of AC magnetic fields, be used to speed up the dynamics? Note that
in present samples, the largest interaction energies are more than
two orders of magnitude above room temperature, so that
further miniaturization is possible without resorting to cryogenics.

Better equilibration might then open the door for an experimental
study of constrained classical \cite{constrclass} and perhaps
eventually even quantum dynamics \cite{reichring} (and quantum ice
\cite{slwc}). Even though this will require a substantial experimental
effort, there appears to be no fundamental obstacle to obtaining at
least a classical ice regime.

The results obtained along the way should provide insights into how
the physics of frustration can lead to new ways of effectively
suppressing interactions between neighboring magnetic nanoislands and
into the limits imposed by disorder, a topic of interest with view to
applications in memory storage \cite{nanoreview}.  We are thus
optimistic that dipolar nanoarrays will provide an interesting field
for further studies.

\suhe{Acknowledgements} We thank P. Schiffer for several very helpful 
explanations and for assistance with the figures, and C. Henley,
A. Middleton, S. Sondhi and O. Tchernyshyov for useful
discussions. Some of our simulations made use of the ALPS library
\cite{ALPS}. This work was in part supported by the Minist\`ere de la
Recherche et des Nouvelles Technologies with an ACI grant.

\end{document}